\documentclass[fleqn,usenatbib]{mnras}

\usepackage{newtxtext,newtxmath,multirow}

\usepackage[T1]{fontenc}

\DeclareRobustCommand{\VAN}[3]{#2}
\let\VANthebibliography\thebibliography
\def\thebibliography{\DeclareRobustCommand{\VAN}[3]{##3}\VANthebibliography}

\usepackage{graphicx}	% Including figure files
\usepackage{amsmath}	% Advanced maths commands

\title[The dual corona in GRO J1655--40]{The X-ray corona in the black-hole binary GRO J1655--40 from the properties of non-harmonically related quasi-periodic oscillations}

\author[Rout et al.]{
Sandeep K. Rout,$^{1}$\thanks{E-mail: skrout@prl.res.in (SKR)}
Mariano M\'endez,$^{2}$
and Federico Garc\'ia, $^{3}$\\
\\
$^{1}$Physical Research Laboratory, Navarangpura, Ahmedabad 380009, Gujarat, India\\
$^{2}$Kapteyn Astronomical Institute, University of Groningen, PO Box 800, NL-9700 AV Groningen, the Netherlands\\
$^{3}$Instituto Argentino de Radioastronom\'a (CCT La Plata, CONICET; CICPBA; UNLP), C.C.5, (1894) Villa Elisa, Buenos Aires, Argentina 
}

\date{Accepted XXX. Received YYY; in original form ZZZ}

\pubyear{2023}

\begin{document}
\label{firstpage}
\pagerange{\pageref{firstpage}--\pageref{lastpage}}
\maketitle

\begin{abstract}

The study of quasi periodic oscillations (QPOs) plays a vital role in understanding the nature and geometry of the Comptonizing medium around black-hole X-ray binaries. The spectral-state dependence of various types of QPOs (namely A, B, \& C) suggests that they could have different origins. The simultaneous presence of different types of QPOs would therefore imply the simultaneous occurrence of different mechanisms. In this work we study the radiative properties of two non-harmonically related QPOs in the black-hole binary GRO J1655--40 detected at the peak of the ultraluminous state during the 2005 outburst of the source. The two QPOs have been previously identified as types B \& C, respectively. We jointly fit the phase-lag and rms spectra of the QPOs and the time-averaged spectrum of the source with the time-dependent Comptonization model {\sc vkompth} to infer the geometry of the media producing the QPOs. The time-averaged spectrum required a hot disk of 2.3 keV and a steep power law with index 2.7, revealing that the source was in an ultraluminous state. The corona that drives the variability of the type-B QPO is smaller in size and has a lower feedback fraction than the one that drives the variability of the type-C QPO. This suggests the simultaneous presence of a horizontally extended corona covering the accretion disk and a vertically elongated jet-like corona that are responsible for the type-C \& B QPOs, respectively. 
 
\end{abstract}


%%%%%%%%%%%%%%%%%%%%%%%%%%%%%%%%%%%%%%%%%%%%%%%%%%

%%%%%%%%%%%%%%%%% BODY OF PAPER %%%%%%%%%%%%%%%%%%

\section{Introduction} \label{sec:intro}

The study of rapid aperiodic variability is a powerful tool in the characterisation of black-hole binaries. The time scales associated with the variability correspond to the motion of matter in the vicinity of the black hole \citep[See reviews by][and references therein]{done07,ingram19}. Quasi periodic oscillations (QPOs), appearing as narrow peaks in the Fourier power spectra of these sources, are excellent tracers of the Comptonizing medium surrounding the black holes \citep[][]{vanderklis89a,nowak00,belloni02}. Low-frequency QPOs ($0.1 - 30$ Hz) are classified into three sub categories, namely types A, B, and C, based on their centroid frequency, strength as well as the shape and variability of the broadband noise \citep{remillard02,casella04}.

Most black-hole binaries display outbursts during which the source goes from being undetected in the so-called quiescence, to a luminosity around the Eddington limit and back to quiescence on timescales of weeks to months \citep{lasota01,frank02}. The ABC classification of QPOs plays an important role in characterising the different states of black-hole binaries \citep[See e.g.,][for a review]{belloni10}. A typical outburst begins in the low/hard state (LHS) where the hard Comptonized component, characterised by a shallow power-law with an index of $\sim 1.6$, dominates the spectrum. The fractional rms amplitude of the variability in the $0.1 - 20$ Hz frequency range is high ($\sim 30 \%$) and the power spectrum contains a type-C QPO \citep{psaltis99,belloni02,pottschmidt03}. As the luminosity increases, the source transitions to the high/soft state (HSS), via the hard (HIMS) and soft intermediate states (SIMS), with corresponding decrease in hardness and broadband variability. The power spectra in the HIMS also show type-C QPOs and the broadband noise have $10\%-20\%$ rms amplitude. In the SIMS type-B or A QPOs are detected and the broadband noise further decreases to a few percent \citep{miyamoto91,belloni05}. While the time-averaged spectra of the two intermediate states are quite similar, with the power-law index lying around $2.0 - 2.4$, the differences become clear in the power spectra \citep{gierlinski05}. The HSS has the highest thermal disk contribution and the lowest variability, $\sim 1\%$ or less \citep{mitsuda84,belloni99}. Sometimes, type-A QPOs are detected during this state \citep{casella04,belloni10,belloni16}. The source stays in the HSS while the intensity decreases and then gradually transitions to the LHS via the same two intermediate states in reverse order. In the decay of an outburst, during the low-luminosity transitions, QPOs are sometimes detected, but less often than during the rise, generally owing to low count rates. Apart from these four standard states, sometimes the source makes an excursion to an anomalous or ultraluminous state (ULS) just after or before the HSS \citep{belloni97,belloni10}. Along with a manifold increase in luminosity, sometimes reaching and/or exceeding the Eddington rate, this state is characterised by low variability ($5\% - 10\%$) and large colour variations \citep{dunn10,uttley15,ingram19}. During the ULS, the time-averaged spectrum remains very soft and the power spectrum shows type-B or C QPOs, thus inviting the moniker ``anomalous" \citep{motta12}. The aforementioned states in black-hole binaries can be easily identified from the hardness-intensity diagram \citep{fender04b} and the absolute rms-intesity diagram \citep{munozdarias11a}. The changes of the source across spectral states are also accompanied by radio emission from a jet or a relativistic outflow \citep{fender04b}. In particular, the transition from the HIMS to the SIMS usually coincides with the launch of relativistic jets that are bright in radio wavebands \citep{russellt19,homan20}. The simultaneous detection of type-B QPOs and radio jets has inspired models explaining the former to be formed by the later \citep{kylafis20}.

With over three decades of extensive timing studies, the phenomenology of QPOs has become fairly clear, although the origin of the QPOs is still debated. The dynamic origin of the oscillations has been broadly ascribed to two categories of models, either due to the intrinsic variability of the accretion disk \citep{tagger99,molteni96,cabanac10,wagoner01}, or to geometrical effects \citep{stella98,ingram09,schnittman06}. These two classes of models are mostly concerned with the dynamic origin of the QPOs, answering basically how a particular frequency in the disk may be excited. While choosing one of the two categories is difficult, model-independent studies suggest a geometrical origin \citep{motta15,vandeneijnden17}. These studies also claim a separate origin for the type-B and C QPOs. \citet{ingram16} carried out phase-resolved spectroscopy of the type-C QPO in H1743--322 that appears to indicate that the components of the reflection spectrum vary with the phase of the QPO, providing evidence for a geometrical origin. However, using the same technique on the type-C QPO in GRS 1915$+$105, \citet{nathan22} found a very small inner disk radius ($\sim$ 1.4 $R_g$) and a very large thermalisation time scale ($\sim$ 70 ms). Such a small inner radius cannot accommodate the necessary inner flow for precession, while the large thermalisation time scale would wash out the short time lags observed in some cases. \citet{stevens16} carried out a similar exercise for the type-B QPO in GX 339--4, and they suggest a possible origin from precession of an extended jet \citep[See also][]{kylafis18,kylafis20}. \citet{mastichiadis22}, on the other hand, showed that QPOs could be formed by a dynamic coupling between the accretion disk and the Comptonizing medium.

While much effort has been directed towards understanding the dynamical origin of the QPO, not much attention has been given to their radiative properties. Out of the few that have tried \citep[e.g.,][]{lee98,lee01,shaposhnikov12,nobili00}, most could explain only the energy-dependent phase-lags but not the rms amplitudes. Recently, \citet{garg20} developed a model incorporating propagation of fluctuations in the components of the time-averaged spectrum to qualitatively explain the time-lag and rms spectra of the QPOs in GRS 1915$+$105. Whatever the process behind the oscillations, it appears to be quite certain that the variability originates from the Comptonizing medium. This becomes evident as the QPOs are mostly detected in hard states, where Comptonization dominates the time-averaged spectrum and, furthermore, the rms of the QPOs generally increases with energy, in some cases reaching about 15\% at 100 keV \citep{bu21}.  

\citet{karpouzas20} built upon the Comptonization scheme of \citet{lee98}, \citet{lee01}, and \citet{kumar14} and developed a model to explain the radiative properties of QPOs in a neutron-star system. This model considers the QPOs to be oscillations of the thermodynamic properties of the accretion flow, such as the temperature and heating rate of the corona, and the seed photon temperature. The model incorporates feedback of a fraction of the  Comptonized photons that return to the seed-photon source. \citet{bellavita22} extended the model by replacing the seed-photon source from a single temperature blackbody, relevant for neutron stars, to a multi-temperature blackbody, suitable for black-hole binaries. Using this model, named \texttt{vkompthdk}, the phase-lag and rms spectra of both type-B and type-C QPOs could be explained \citep{garciaf21,zhangy22,zhangy23,peirano23,rawat23}. The results from these studies suggest that the geometry of the Comptonizing medium is different for the two types of QPOs. While in the case of type-B QPOs the corona has a vertically  elongated (jet-like) geometry, the corona for type-C QPOs is more horizontally extended and significantly covers the accretion disk.

The two types of QPOs, B and C, mostly appear in separate states, it is only in rare cases that they appear together. If their origin are indeed from separate structures, as found earlier, simultaneous type-B and C QPOs would imply a complex structure of the corona with both vertically elongated and horizontally extended components. To test this hypothesis, we study the simultaneous type-B and C QPOs detected during the ULS of GRO J1655--40 (henceforth, J1655) during its 2005 outburst \citep{debnath08,motta12}. This outburst was densely monitored by the Rossi X-ray Timing Explorer (RXTE), showing that the source traced all spectral states. Following the HSS, for about twenty days, J1655 made an excursion to the ULS that was apparent in the hardness-intensity diagram as a huge increase in count rate along with marginal increase in hardness and variability. The power spectrum during this state consisted of a type-C QPO along with a broad peak with a similar frequency-rms relation as that for type-B QPOs \citep{motta12}. When the source reached the peak luminosity within the ULS (and also overall), the broad power-spectral peak transitioned into a proper type-B QPO. The observation during that period, dated 2005 May 18 (ObsID 91702015800), consisting of both a type-B and C QPO is analysed here. In Section \ref{sec:analysis} we describe the data reduction and analysis methods. In Section \ref{sec:results} we present the main result from the simultaneous fits to the time-averaged spectrum of the source and the rms and phase-lag spectra of the QPOs. Finally, in Section \ref{sec:discussion} we discuss the implications of the result of our analysis on the origin of the QPOs and the geometry of the Comptonizing medium.

\section{Data Reduction and Analysis} \label{sec:analysis}

We reduced the Proportional Counter Array (PCA) binned data using the new tools in \texttt{heasoft-6.31}. The new tools are a wrapper of the old extraction tools making the analysis simpler. We extracted the source spectrum in the $3 - 30$ keV range from the \texttt{Standard 2} mode using all standard filtering criteria. We selected the data from all three xenon layers of the Proportional Counter Unit (PCU) 2 for the source spectrum. We computed the background spectrum using the bright background model \texttt{pca\_bkgd\_cmvle\_eMv20111129.mdl}.

We carried out the timing analysis with the \texttt{IDL}-based timing package \texttt{GHATS} \footnote{http://astrosat.iucaa.in/$\sim$astrosat/GHATS\_Package/Home.html}. The high time resolution PCA data were packaged in two single bit modes with 125 $\mu$s resolution and one event mode with 62 $\mu$s resolution. The single bit mode data were binned into two broad channel ranges, one spanning channels 0 to 13 ($2.06 - 7.35$ keV) and the other covering channels 14 to 35 ($7.46 - 16.43$ keV). The event mode, on the other hand, consisted of 32 channels from 36 to 249. We constructed the broadband power spectrum by averaging over the full $3 - 30$ keV lightcurve segments of 16 s and a time resolution of 1/4096 s giving a Nyquist frequency of 2048 Hz. We removed the Poisson noise by subtracting the average power between $1000 - 2000$ Hz from the Leahy normalized power spectrum. We then converted the normalization to squared fractional rms following \citet{belloni90}. We fitted the power spectrum by a multi-Lorentzian model in \texttt{xspec} \citep{belloni02}. The model required five \texttt{lorentz} components to fit the $0.01 - 100$ Hz spectrum. Out of these Lorentzians, three fitted the broadband noise and two were required for the QPOs. The lower frequency QPO at 6.8 Hz is the type-B QPO and the higher frequency QPO at 19.0 Hz is the type-C QPO \citep{motta12}. 

\begin{figure*}
    \centering
    \includegraphics[scale=0.6]{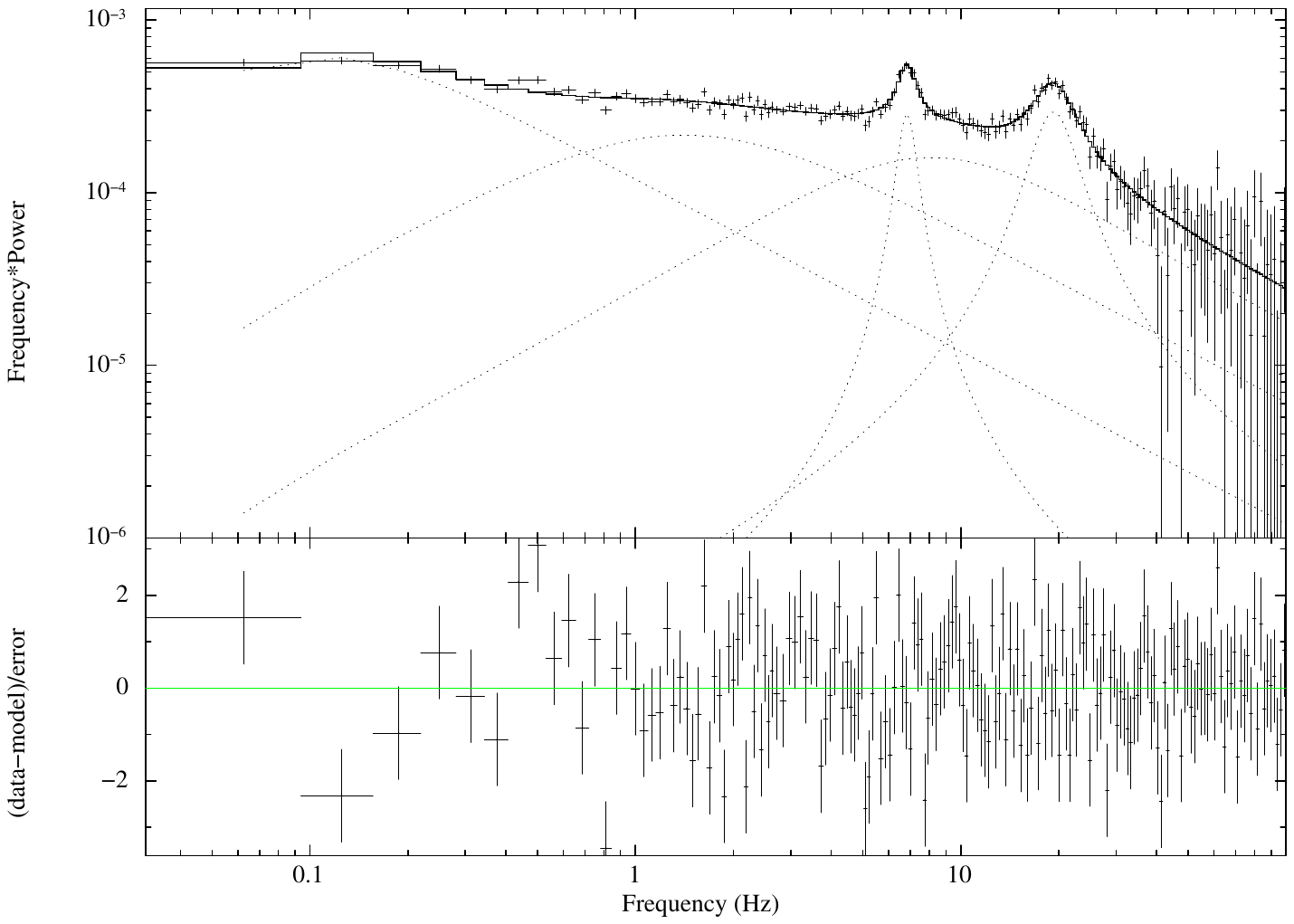}
    \caption{Top panel: The power spectrum of GRO J1655--40 fitted with a multi-Lorentzian model. The dotted lines are the individual Lorentzian components used in the fitting. The solid line represents the total model. The peak at $\approx 7$ Hz is the type-B QPO and the peak at $\approx 20$ Hz is the type-C QPO. Bottom panel shows the residuals of the fit.}
    \label{fig:pds}
\end{figure*}

We constructed the rms spectra for each of the two QPOs by extracting the power spectra for smaller energy bins and fitting them with the same multi-Lorentzian model that we used for the full band. While computing the rms we fixed the centroid frequency and the full width at half maximum (FWHM) to the values obtained from fitting the full band. It was possible to make significant measurement for six energy bins in the $2 - 30$ keV range. The common way to compute the phase lags of QPOs is to average the real and imaginary part of the cross-spectrum in a frequency range dominated by the QPO signal, usually the FWHM of a QPO. This works well when the QPO is very strong and not contaminated by other components. The two QPOs in the power spectrum of J1655 are quite close to each other and their respective Lorentzians have significant overlap, both with each other and with Lorentzians of the adjacent continuum (Figure \ref{fig:pds}). Therefore, we use a novel technique to extract the phase-lags from the entire region of the power spectrum where the QPO is present \citep{peirano22,alabarta22}. We constructed the cross spectra using the full energy band as the reference band and the six smaller energy bins used for the rms spectra as the subject bands. We jointly fitted the real and imaginary parts of the cross spectra with a multi-Lorenzian model. We also fitted a constant to the real part of the cross-spectrum to account for the cross-correlation introduced from using the full band as reference. Apart from that we introduced an additive model using \texttt{mdefine} with a single parameter for capturing the phase-lag ($\Delta \Phi$). We fixed the frequency and FWHM of the Lorentzians to the values obtained from fits to the full energy power spectrum. The normalization of the real part was left free to vary while that of the imaginary part was linked to the phase-lag as $\operatorname{Im}(C) = \operatorname{Re}(C) \times \tan{(\Delta \Phi)}$ where $C$ is the complex cross-spectrum. In this way the phase-lag is obtained as an \texttt{xspec} parameter and takes into account the entire profile of the QPO in the full frequency range. 

With all the products ready we jointly fitted the fractional rms and phase-lag spectra of both QPOs along with the time-averaged energy spectrum of the source. The time-averaged spectrum required the typical combination of an absorbed (\texttt{TBabs}) disk \citep[\texttt{diskbb};][]{mitsuda84}, Comptonization \citep[\texttt{nthComp};][]{zdziarski96,zycki99} and reflection \citep[\texttt{relxillCp};][]{dauser14,garcia14} components. We adopted the abundance in the \texttt{TBabs} component from \citet{wilms00} and set the cross-sections according to \citet{verner96}. We fixed the neutral H column density to $7.5 \times 10^{21}$ cm$^{-2}$ \citep{diaztrigo07, brocksopp06}. The \texttt{diskbb} component represents the emission of a multi-temperature blackbody based on the $\alpha$-disk prescription of \citet{shakura73} and is parameterised by the inner-disk temperature ($kT_{in}$) and a normalization ($N_{dbb}$). The Comptonization component \texttt{nthComp} is approximately equal to a power law with index $\Gamma$, along with a high energy cutoff and a low energy rollover. The high energy cutoff is parameterised by the electron temperature, $kT_e$, and the low-energy rollover by the seed photon temperature, $kT_{bb}$, which in our case was tied to $kT_{in}$ from \texttt{diskbb}. \texttt{nthComp} has a switch $inp\_type$ to toggle between a single temperature blackbody relevant for neutron star binaries and a disk blackbody pertinent for black hole binaries. We chose the later. \texttt{relxillCp} is a relativistic reflection model with the irradiating flux, or emissivity, assumed to be a broken power law. The outer emissivity index was fixed to the Newtonian value of 3 and the break radius was fixed at 6 $R_g$. The inner emissivity index, $q1$, was allowed to vary within $3 - 10$. The primary source spectrum is the same as \texttt{nthComp}, therefore, we tied $\Gamma$ and $kT_e$ parameters in \texttt{relxillCp} to the respective parameters in \texttt{nthComp}. Since we had separately included a Comptonization component, the $refl\_frac$ parameter was fixed at -1 letting \texttt{relxillCp} to contribute only the reflection part. We fixed the Fe abundance at solar values and left the disk density and ionization parameters free to vary. 

We fitted the phase-lag and rms spectra of the QPOs with the time-dependent Comptonization model \texttt{vkompthdk} \citep{bellavita22}. This variant incorporates \texttt{diskbb} as the seed-photon source and considers a single corona. The time-averaged version of \texttt{vkompthdk} is essentially the same as \texttt{nthComp}. We tied the seed-photon temperature, power-law index, and coronal temperature in \texttt{vkompthdk} to the $kT_{in}$ of \texttt{diskbb} and $\Gamma$ and $kT_e$ of \texttt{nthComp}, respectively. The model is broadly insensitive to the $af$ parameter. We verified that by tying $af$ to the inner disk radius obtained from the normalization of \texttt{diskbb} \citep{makishima86,kubota98}. The fit statistics and the best-fitting parameters remained unaffected. Therefore, we fixed $af$ to the default value of 250 km. Apart from these, \texttt{vkompthdk} has the coronal size, $L$, the feedback fraction, $\eta$, the variability in the external heating rate of the corona, $\delta \dot{H}_{\rm ext}$, and an additive reference lag, $reflag$, as free parameters. $reflag$ is the lag of the model in the $2-3$ keV band and adjusts the lag to account for the fact that the lag spectrum is defined except for an additive constant. This allows the user to choose the reference band arbitrarily. The feedback fraction, $0 \leq \eta \leq 1$, is given as $$\eta = \frac{F_{\rm D,f} - F_{\rm D,o}}{F_{\rm D,f}}$$ where $F_{\rm D,f}$ is the flux of the disk after it has thermalised the irradiated hard photons from the corona and $F_{\rm D,o}$ is the flux of the disk before the corona illuminates it. Thus, $\eta$ represents the fraction of the total disk flux that arises due to the feedback of coronal photons. Internally, however, the model calculates an intrinsic feedback fraction, $\eta_{\rm int}$, which is the fraction of Comptonized photons that illuminates the disk \citep{karpouzas20}. The $\eta_{\rm int}$ is not a free parameter in the model as its maximum attainable value changes with the disk and the coronal parameters, which makes it impractical to use it in \texttt{xspec}.  

\begin{figure*}
    \centering
    \includegraphics[scale=.6]{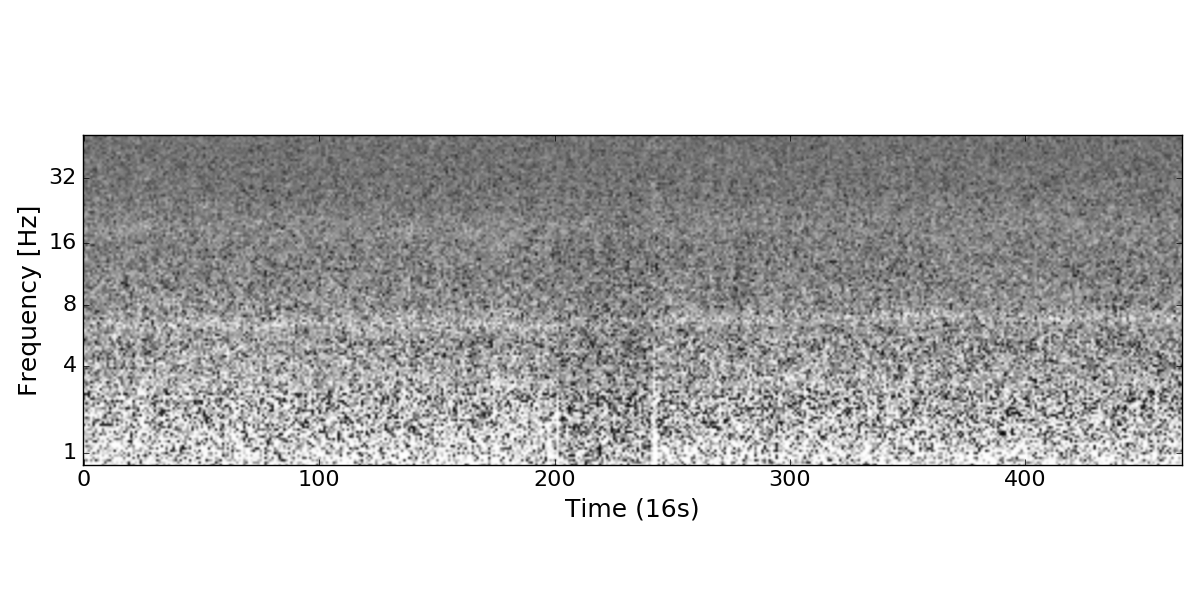}
    \caption{Spectrogram for the observation showing the two simultaneous QPOs at $6.8$ Hz and $19$ Hz. Each bin in the time axis spans 16 seconds.  }
    \label{fig:dynpds}
\end{figure*}

Owing to the low resolution of the PCA spectrum, the spin and inclination parameters in \texttt{relxillCp} cannot be constrained. Therefore we strove to fix these parameters to the best estimates from the literature. However, there are multiple contradicting reports making the choice tricky. By modeling the three simultaneous QPOs with the relativistic precession model, \citet{motta14a} and \citet{rink22} arrived at a low spin $a \sim 0.3$. \citet{shafee06} reported a spin $a = 0.7$ from continuum fitting, while \citet{reis09} measured $a > 0.9$ using reflection spectroscopy. \citet{reis09} noted that the spin will be greater than 0.9 using continuum fitting after using the correct distance. A high spin was further suggested by \citet{diaztrigo07} based on reflection fits with a relatively high inclination. \citet{stuchlik16} attempted to address the conflicting spin estimates and found that by using an epicyclic resonance model, and abandoning the assumption that all QPOs originate at a single radius, the spin from QPOs can become consistent with the Fe line results. Deciding the true inner disk inclination is even more complicated. While the orbital inclination was found to be 70$^\circ$ \citep{greene01}, the jet was reported to be tilted at 85$^\circ$ \citep{hjellming95} suggesting spin-orbit misalignment. However, with reflection spectroscopy the inclination was found to be 30$^\circ$ \citep{reis09} or 52$^\circ$ \& 63$^\circ$ \citep{diaztrigo07}. With such a variety of options it is only meaningful to try different combinations and see how the data respond. 

One important aspect to be dealt with is that the \texttt{vkompthdk} model is a purely Comptonization model whereas the time-averaged spectrum consists of significant contribution from the disk and reflection components. Therefore, the rms calculated by the model is only from the Comptonized component, whereas the observed rms will be diluted by the effects of the disk and reflection, which are assumed to be much less variable than the corona. To take into account this effect we multiplied a \texttt{dilution} component to \texttt{vkompthdk} only for the rms spectra. The dilution is defined as the fraction of the total flux contributed by the \texttt{nthComp} component, \texttt{nthComp}$(E)$/(\texttt{diskbb}$(E)$ + \texttt{nthComp}$(E)$ + \texttt{relxillCp}$(E)$)  \citep{bellavita22}.

  \begin{figure*}
    \centering
    \includegraphics[scale=0.5]{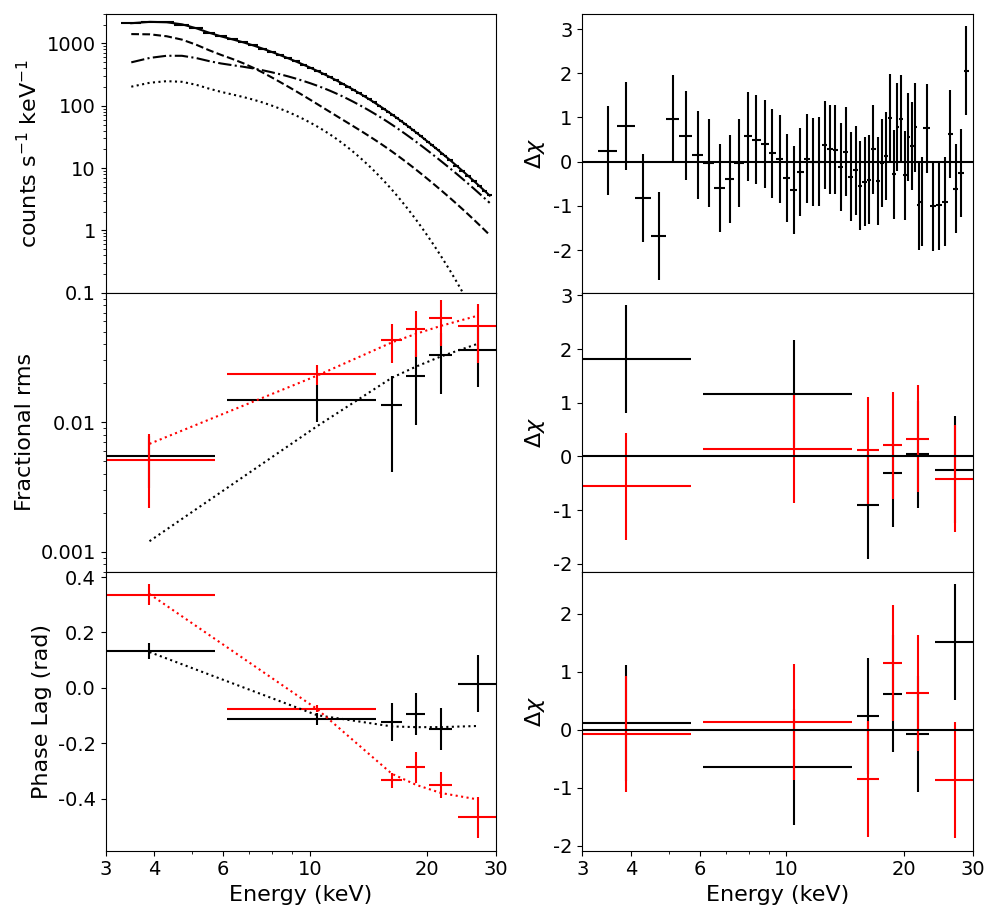}
    \caption{From top to bottom, the left panels represent, respectively, the time-averaged spectrum of the source and the fractional rms and phase-lag spectra of the QPOs. The corresponding panel in the right column are the residuals. The black and red points in the middle and bottom panels represent the type-B and C QPO, respectively. The dotted lines are the best-fitting models. In the top left panel, the dotted, dashed, and dot-dashed lines represent the disk, Comptonization, and reflection components, respectively.}
    \label{fig:datamod}
\end{figure*}

\section{Results} \label{sec:results}

We show the $0.03 - 100$ Hz power spectrum of J1655 in Figure \ref{fig:pds}. The power spectrum has a flat top spanning about three decades in frequency and falls off at around 20 Hz. It can be decomposed into five Lorentzians and has a broadband $0.03 - 100$ Hz fractional rms of $5.4\%$. Out of the five components, three Lorentzians account for the broad continuum and the other two fit the QPOs at 6.8 Hz and 19.0 Hz. The former is a type-B QPO and the later is a type-C QPO having fractional rms amplitudes of 0.8\% and 1.2\%, respectively. The simultaneity of the two QPOs is demonstrated by the spectrogram shown in Figure \ref{fig:dynpds}. This classification is based on the fact that the two QPOs lie on the two branches corresponding to the types-B and C on the broadband rms - QPO frequency plane and the type-B QPO at 6.8 Hz is highly variable in frequency \citep[See Figure 5 in][]{motta12}. While this classification helps in contextualizing our work within the existing literature, it is worth noting that our conclusion is independent of the QPO classification.  

Deciphering the radiative origin of these two QPOs is the objective of this work. The time-averaged spectrum of the source and the fractional rms and phase-lag spectra of the QPOs were jointly fitted in \texttt{xspec}. As explained in detail in section \ref{sec:analysis}, the PCA spectrum could not discriminate between the low (0.3) and high (0.9) spin values, yielding an acceptable fit in both cases. However, the inclination seemed to be correlated to the spin, with a lower spin giving a low inclination ($\sim 20^\circ$) and a higher spin leading to a high inclination ($\sim 50^\circ$) when left free. Trying to freeze the inclination to the orbital tilt ($70^\circ$) or the jet orientation ($85^\circ$) did not result in acceptable fits. Instead, leaving the spin free with the inclination fixed at $70^\circ$ gave an acceptable fit, albeit with a much higher spin estimate of $\sim 0.99$. It is worthwhile to remember that reflection spectroscopy with XMM-Newton data also resulted in very similar spin and inclination estimates \citep{reis09, diaztrigo07}. 

 \begin{table*}
\centering
\caption{The best-fitting parameters from the simultaneous fit to the time-averaged spectrum of the source and the rms and lag spectra of the QPOs. The three models correspond to the three choices of spin and inclination as mentioned in the top row. The $\eta_{\rm int}$ is not a model parameter and it is computed separately. The errors represent the 1$\sigma$ uncertainty.}

\label{tab:bfp}
\begin{tabular}{||c c|c c c||}

 &  & Model 1 & Model 2 & Model 3 \\
Component & Parameter & $a = 0.3$ & $a = 0.9$ & $a$ free  \\
 & & $i$ free & $i$ free & $i = 70^\circ$ \\
 \hline \hline

 \multirow{2}{*}{\texttt{diskbb}} & $kT_{in}$ (keV) & $2.33 \pm 0.04$ & $2.43 \pm 0.04 $ & $2.44 \pm 0.03$ \\
 & $norm$ & $23.3 \pm 1.4$ & $24.6 \pm 1.3$ & $32.2^{+3.2}_{-2.6}$ \\

 \hline

\multirow{3}{*}{\texttt{nthComp}} & $\Gamma$ & $2.74 \pm 0.01$ & $2.74 \pm 0.01 $ & $2.76 \pm 0.01$ \\
& $kT_e$ (keV) & $160^{+46}_{-24}$ & $> 190 $ & $>202$ \\
& $norm$ & $3.45 \pm 0.18$ & $2.97 \pm 0.19 $ & $2.9 \pm 0.1$ \\

\hline

\multirow{6}{*}{\texttt{relxillCp}} & $i$ ($^\circ$) & $18.9 \pm 3.1$ & $51.6 \pm 1.4 $ & - \\
 & $a$ & -  & -  & $0.990 \pm 0.001$ \\
& $q1$ & $3.5 \pm 0.2$ & $7.8^{+1.1}_{-0.8}$ & $>9.7$ \\
& $\log \xi$ (erg cm s$^{-1}$) & $4.5 \pm 0.1$ & $4.3 \pm 0.1 $  & $3.5 \pm 0.1$ \\
& $\log N$ (cm$^{-3}$) & $16.5 \pm 0.1$ & $16.4 \pm 0.1 $ & $16.7 \pm 0.1$ \\
& $norm$ & $0.72 \pm 0.03$ & $1.30 \pm 0.05 $ & $2.7 \pm 0.1$ \\

\hline

 \multirow{5}{*}{\texttt{vkompthdk}} & $L$ (km) & $963^{+153}_{-128}$ & $849^{+157}_{-109}$ & $660^{+92}_{-67}$ \\
 \multirow{5}{*}{(Type-B QPO)} & $\eta$ & $0.20 \pm 0.03$ & $0.21 \pm 0.03 $ & $0.15 \pm 0.01$ \\
 & $\eta_{\rm int}$ & $0.041 \pm 0.003$ & $0.057 \pm 0.009$ & $0.057 \pm 0.007$\\
& $\delta \dot{H}_{\rm ext}$ & $0.05 \pm 0.01$ & $0.06 \pm 0.01 $ & $0.07 \pm 0.01$ \\
& $reflag$ & $0.32 \pm 0.06$ & $0.35^{+0.06}_{-0.09} $ & $0.98^{+0.32}_{-0.36}$ \\

\hline

 \multirow{5}{*}{\texttt{vkompthdk}} & $L$ (km) & $1160^{+69}_{-74}$ & $1134^{+72}_{-69} $ & $1135^{+59}_{-52}$ \\
  \multirow{5}{*}{(Type-C QPO)} & $\eta$ & $0.60 \pm 0.05$ & $0.57 \pm 0.05 $ & $0.58 \pm 0.04$ \\
& $\eta_{\rm int}$ & $0.157 \pm 0.008$ & $0.16 \pm 0.01$ & $0.15 \pm 0.02$ \\
 & $\delta \dot{H}_{\rm ext}$ & $0.09 \pm 0.01$ & $0.09 \pm 0.01 $ & $0.10 \pm 0.01$ \\
& $reflag$ & $0.44 \pm 0.04$ & $0.46 \pm 0.04 $ & $0.45 \pm 0.03$ \\

\hline \hline
 \multicolumn{2}{c}{$\chi^2$/dof} & 38.87/60 & 35.50/60 & 42.99/60 \\

\end{tabular}
\end{table*}

Table \ref{tab:bfp} gives the best-fitting parameters with three variations of the model by fixing the spin and inclination to different values. Two combinations involve fixing the spin to 0.3 and 0.9 while keeping the inclination free, while the third combination keeps the inclination fixed at $70^\circ$ and allows the spin to vary. The errors quoted represent $1\sigma$ uncertainty and were evaluated by running Markov Chain Monte Carlo simulations in \texttt{xspec} with 1000 walkers and $2\times 10^5$ steps with an initial burn-in length of 40000 steps to ensure the convergence of the chains. We re-scaled the covariance matrix by $10^{-4}$ in order to sample a wide parameter space by the walkers. It is apparent that most of the best fitting parameters remain consistent across the three variants of the models. The only significant differences are the fluxes of the individual components of Model 3 ($i=70^\circ$) which are different from those of Models 1 \& 2. Apart from these the inner disk temperature and disk ionization parameter are, respectively, slightly higher and lower in Model 3 than in Models 1 \& 2. The high disk temperature  $kT_{in}$ $\sim 2.4$ keV and photon index, $\Gamma \sim 2.7$, confirms the ultra soft nature of the source during the observation. The electron temperature could not be constrained for Models 2 \& 3, pegging at the upper limit of 250 keV. For model 1, $kT_e$ was constrained to $\approx 160$ keV. From the reflection component, we found that both the inner emissivity power-law index ($\gtrsim 3$) and disk ionization ($\gtrsim 10^{3.5}$ erg cm s$^{-1}$) remained high. The disk density, on the other hand, was constrained to moderate values of $N \sim 10^{16.5}$ cm$^{-3}$. Figure \ref{fig:datamod} shows the fitted time-averaged spectrum of the source and the rms and phase-lag spectra of the QPOs for the Model 2.

\begin{figure*}
    \centering
    \includegraphics[scale=1]{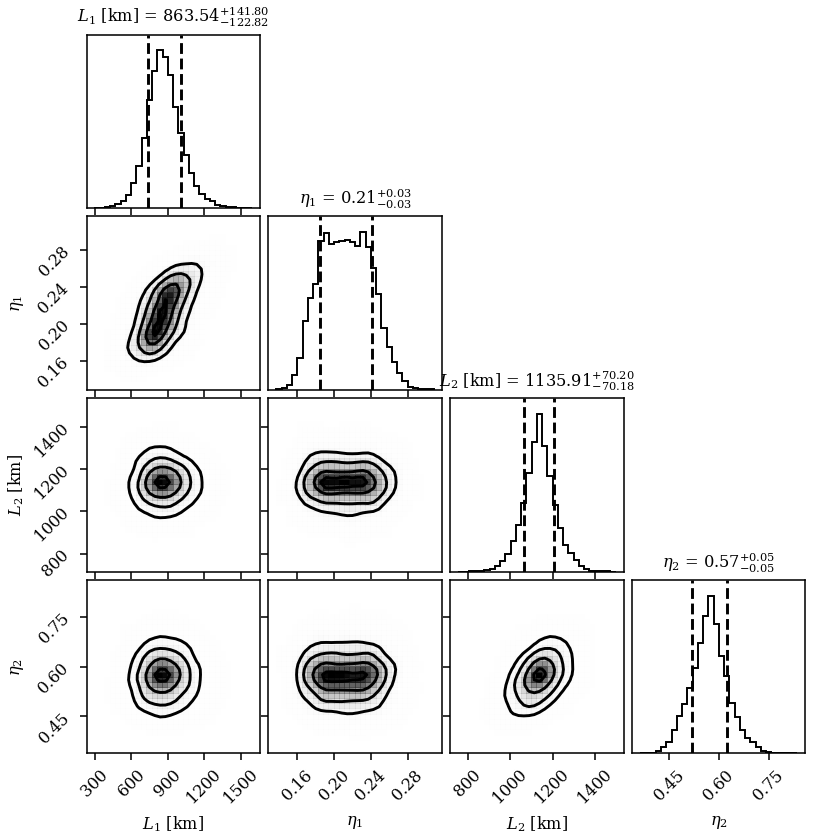}
    \caption{The confidence contours of the geometry parameters (size, $L$, and feedback fraction, $\eta_1$) from the best fit with the \texttt{vkompthdk} model-2. The 1 and 2 subscripts represent the type-B and C QPO, respectively. The corner plot is generated using the tool \texttt{pyXspecCorner}.}
    \label{fig:tcorner}
\end{figure*}

The remarkable result of the joint fitting exercise is that the parameters pertaining to the coronal geometry from \texttt{vkompthdk} remained consistent across the three model variants. The corona responsible for the type-B QPO has a size of $\approx 820$ km and a low feedback fraction of $\sim 20\%$. For the type-C QPO, on the other hand, the corona is much larger, with a size of $\approx 1150$ km and the feedback fraction is also high at $\sim 60\%$. The corresponding intrinsic feedback fractions, $\eta_{\rm int}$, are $\sim 5\%$ and $\sim 16\%$ for the type-B and C QPOs, respectively. The external heating rate for both QPOs is similar, varying between $0.05 - 0.1$. Finally, the reduced $\chi^2$ indicate that all three models provide excellent fits and it is not possible to choose one over the other. We can conclude that a particular choice of the spin and inclination does not drastically affect the joint fits. Most importantly, the geometry of the corona obtained from \texttt{vkompthdk} remains completely unaffected. In Figure \ref{fig:tcorner} we show the confidence contours of the relevant coronal parameters (i.e., size and the feedback fraction) obtained by carrying out Monte Carlo simulations for Model 2. The corresponding contours from Mode1 1 \& 3 are almost identical to those of Model 2. The contours for $\eta_1$ appears to be bimodal, however, the separation of the two lobes is merely at a few percent level.

\section{Discussion} \label{sec:discussion}

The power spectrum of J1655 at the peak of the ultraluminous state during its 2005 outburst displays two non-harmonic QPOs (Figure \ref{fig:pds}), which were identified as being types-B \& C \citep{motta12}. We jointly fitted the time-averaged spectrum of the source and the phase-lag and rms spectra of the QPOs to infer the radiative origin of the QPOs. The phase-lag and rms spectra were modeled with the recently developed time-dependent Comptonization  model \texttt{vkompthdk} \citep{karpouzas20,bellavita22}. The type-B QPO required a corona of size $\approx 820$ km and an intrinsic feedback of $\sim 5\%$. The type-C QPO, on the other hand, required a coronal size of $\approx 1150$ km and a higher feedback of $\sim 16\%$. Using only spectral-timing methods, we show for the first time the co-existence of a vertically elongated, or jet-like corona, and a horizontally extended, or slab-like corona, the oscillations of which result in the type-B and C QPOs, respectively.

Hard X-rays in black-hole binaries are produced by inverse Compton scattering of disk photons in an optically thin cloud of electrons known as the corona \citep{sunyaev79}. Since the variability amplitude increases with energy, \citet{sobolewska06} posited that the radiative properties of the power spectral components should have their origin in the corona. The most prominent of these variability components are QPOs, which appear as narrow peaks in the power spectrum. The QPOs are classified into types A, B, \& C based on their frequency and other variability properties (See Section \ref{sec:intro}). The different types of QPOs have different lag/rms spectra and thus believed to originate from different parts of the corona or a corona with properties that change as the source moves through different states. The increase in rms with energy for both the QPOs (Figure \ref{fig:datamod}), however, suggests a common process likely connected with inverse Comptonization. The same process also gives rise to hard time lags since the hard photons, having undergone Comptonization, arrive later than the soft seed photons \citep{miyamoto88}. Sometimes a fraction of these scattered hard photons travels back and gets thermalised in the accretion disk. Whenever this newly thermalised photons dominate over the intrinsic emission of the disk, the soft photons will arrive later than the hard photons resulting in a soft lag \citep{uttley11, uttley14, demarco15}.

The appearance of the QPOs in the ULS indicates a peculiar change in the accretion flow. In the state transition scheme of black hole binaries, the ULS is both preceded and followed by the HSS. During the HSS, the corona likely disappears indicated by an increase in the fraction of the disk flux, the disappearance of QPOs, and a decrease in the total fractional variability. The re-appearance of the QPOs during the ULS suggests the restoration of the corona for a short period. It is noteworthy that the two QPOs simultaneously occurring suggests a common emission mechanism for the two, despite different dynamical origins. This is precisely the assumption in the \texttt{vkompthdk} model where all kinds of QPOs are generated by the same process of inverse Comptonization. The fact that the lags are different for the two QPOs (See Figure \ref{fig:datamod}) suggests that the photons travel through regions with different geometrical properties, which makes their radiative properties different.

The \texttt{vkompthdk} model used in this work includes a feedback loop akin to the physical scenario where a fraction of the hard coronal photons return to the disk and are re-emitted, accounting for the soft lags. The fraction of the primary hard emission that returns to the disk, which is parameterized in the model as $\eta$, can be used as a proxy for the geometry of the Comptonizing medium \citep{mendez22,garciaf22}. A higher feedback fraction implies that the disk is enveloped by the Comptonizing medium, increasing the chance of back scattered photons to impinge onto the disk. This can happen when a horizontally extended corona covers a significant part of the disk. Accordingly, a lower feedback fraction would mean that the corona and the disk are spatially separated in such a way that the back scattered photons find it difficult to reach the disk. A vertically elongated or jet-like geometry of the corona could explain the low feedback fraction. The \texttt{vkompthdk} model assumes a spherical and homogeneous corona of uniform density and electron temperature. Despite this simplified picture, which are necessary for solving the equations, it is remarkable that the model is able to explain the radiative properties of different types of QPOs.

\begin{figure*}
    \centering
    \includegraphics[scale=.4]{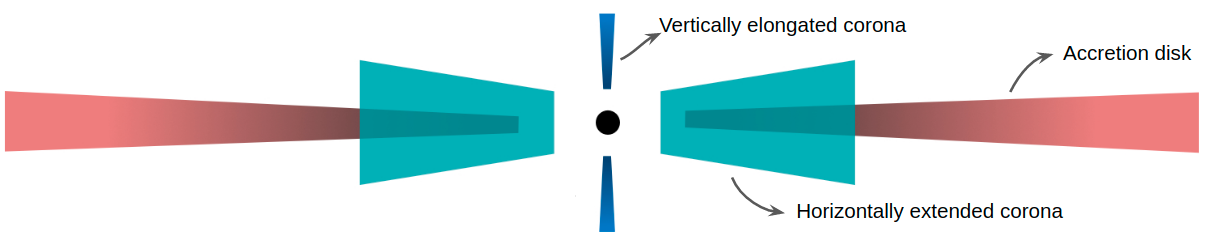}
    \caption{A schematic of the geometry of the two coronae simultaneously present during the observation. Comptonization in the vertically elongated jet-like corona resulted in the type-B QPO while Comptonization in the horizontally extended corona led to the type-C QPO.}
    \label{fig:toy}
\end{figure*}

The size of the corona and the feedback fraction are mainly constrained by the phase-lag spectra of the QPOs. While the amplitude of the lags constrains the size, the relative sign of the lags constrains the feedback fraction. This is intuitive, as a larger corona would imply that the seed photons spend more time inside the corona resulting in longer lags. Accordingly, a negative lag implies that the Comptonized photons scatter back and impinge the disk which re-emits thermally after the correlated coronal emission. This can happen when the corona physically envelopes the disk. Conversely, a positive lag indicates a geometry where the corona is physically detached, or far away, from the disk so that feedback is difficult to manifest. The phase-lag spectra of the two QPOs in J1655 (Figure \ref{fig:datamod}) clearly discerns these two scenarios. The phase-lag spectrum of the type-B QPO is flat and has a low amplitude, whereas the phase-lag spectrum of the type-C QPO is inverted, with negative lags increasing with energy. This suggests that the type-B QPO originates from a smaller corona which is detached from the disk and the type-C QPO arises from a bigger corona which envelopes the disk so as to facilitate higher feedback. The best-fit parameters from our fits with the \texttt{vkompthdk} model correspond quite well with this geometry (See Table \ref{tab:bfp}). The type-B QPO is produced by a corona of $\approx 820$ km and an intrinsic feedback fraction of $\sim 5\%$ while the type-C QPO comes from a larger corona of $\approx 1150$ km and a relatively higher intrinsic feedback of $\sim 16\%$. This would mean the simultaneous presence of two different coronal geometries during the observation. The type-C QPO originates from the oscillations in a larger, horizontally extended, corona while the type-B QPO arises from oscillations in a smaller, vertically elongated, or jet-like, corona. Figure \ref{fig:toy} represents schematically the scenario described here, where two physically distinct Comptonizing media are simultaneously present. The \texttt{vkompthdk} model used here is oblivious to the physical origin of the dynamics of the QPOs. It only explains the radiative origin by assuming the coronal and disk properties to oscillate at the QPO frequency. The oscillations could arise either due to geometrical factors or intrinsic variability (Section \ref{sec:intro}). \citet{mastichiadis22} recently showed that the QPOs could be dynamically produced by a resonant coupling between the accretion disk and the hot Comptonizing corona, which is in agreement with the assumptions of \texttt{vkompthdk}. Although the reflection component was significant, using the PCA spectrum we were not able to constrain the spin of the black hole and the inclination of the disk by fitting the relativistic Fe K$\alpha$ line. We verified that fixing the spin parameter to 0.3 \citep{motta14a} or 0.9 \citep{reis09,stuchlik16} while allowing the inclination to vary, or fixing the inclination to $70^\circ$ \citep{greene01} while varying the spin gives identical fits. Not only are the three variants statistically similar, but the best-fitting parameters are also consistent with being the same (Table \ref{tab:bfp}).

There is increasing evidence that the type-B QPOs are associated with the jet emission \citep{fender09,homan20,garciaf21}. The transition from the HIMS to the SIMS is marked by both the appearance of a type-B QPO and ejection of a relativistic jet \citep{fender04b,russellt19,homan20}. The type-B QPO variability has also been seen to correlate with the system inclination, suggesting a jet origin \citep{motta15,reig19}. \citet{kylafis20} were able to explain the variability of the photon index of the Comptonizing component by considering the Comptonization to take place in a precessing jet. On the other hand, the type-C QPOs have been shown to originate from an extended corona. For instance, the geometrical model for the formation of type-C QPOs considers the oscillations to arise from a precessing inner accretion flow \citep{ingram09}. Recently, polarization measurement of Cyg X-1 in the hard state indicated a horizontally extended, slab-like, corona \citep{krawczynski22}. Since during the hard states we generally detect type-C QPOs, we can associate them with a horizontally extended corona, akin to what we got for J1655. 

Previous studies using the \texttt{vkompthdk} model (and its dual-component variant \texttt{vkdualdk}) are broadly consistent with the aforementioned idea. The evolution of the type-C QPOs during the HIMS in MAXI J1535--571 shows that the corona size initially decreases and then increases rapidly towards the end of the HIMS. The feedback fraction, on the other hand, evolves conversely suggesting that the corona transformed from a horizontal to a vertical structure \citep{rawat23,zhangy22}. Towards the end of the HIMS, the corona of MAXI J1535--571 reached its maximum size. The transition from the HIMS to the SIMS was coincidental with the ejection of a transient jet and the appearance of type-B QPOs. \citet{zhangy23} analyzed these type-B QPOs and showed that the corona contracted in the vertical direction, and the feedback fraction gradually increased along the SIMS. The type-B QPOs in MAXI J1348--630 and GX 339--4 have been shown to require two physically connected structures \citep{garciaf21,bellavita22,peirano23}. Often the larger of the two structures is interpreted as a jet and the smaller structure, with a higher feedback fraction, is associated with a compact corona. These studies with the dual-component model (\texttt{vkdualdk}) suggest that the corona harbours a complex geometry with the presence of both a horizontal and vertical structure. Depending on the strength of their variability, one or both the structures manifest themselves in the data. Recently, \citet{mendez22} found a strong connection between the corona and the jet in GRS 1915$+$105, suggesting that the former morphs into the later due to a redistribution of the accretion power. While they observed only type-C QPOs during the HIMS, the changes in the corona could be due to the source approaching the SIMS. Since type-B QPOs were not detected, the source probably never made the transition to the SIMS. Our analysis of J1655 during the ULS demonstrates one such rare instance when the transition has taken place and both the corona and the jet are simultaneously present.

% \section{Summary} \label{sec:summary}

\section*{Acknowledgements}

SKR acknowledges the support of the COSPAR fellowship programme for partially funding a visit to the University of Groningen which culminated in this research. SKR also acknowledges the support of Physical Research Laboratory which is funded by Department of Space, Government of India. MM and FG have benefited from discussions during Team Meetings of the International Space Science Institute (Bern), whose support they acknowledge. MM and FG acknowledge the research programme Athena with project number 184.034.002, which is (partly) financed by the Dutch Research Council (NWO).  FG is a CONICET researcher and acknowledges support by PIBAA 1275 and PIP 0113 (CONICET). 

%%%%%%%%%%%%%%%%%%%%%%%%%%%%%%%%%%%%%%%%%%%%%%%%%%
\section*{Data Availability}

The X-ray data used in this article are accessible at NASA’s High Energy Astrophysics Science Archive Research Center https://heasarc.gsfc.nasa.gov/. The time-dependent Comptonization model and the generator of the MCMC corner plot are available at the GitHub repositories https://github.com/candebellavita/vkompth and https://github.com/garciafederico/pyXspecCorner.



\bibliographystyle{mnras}
\bibliography{references} 

%%%%%%%%%%%%%%%%%%%%%%%%%%%%%%%%%%%%%%%%%%%%%%%%%%

% Don't change these lines
\bsp	% typesetting comment
\label{lastpage}
\end{document}